\documentclass[aps, twocolumn]{revtex4-1}
\usepackage{adjustbox}
\usepackage{epsfig}
\usepackage{graphicx}
\usepackage{amsmath,amssymb,amsfonts}
\usepackage{array}
\usepackage{url}
\usepackage{soul}
\usepackage[colorlinks=true,linktocpage=true,linkcolor=blue,citecolor=blue]{hyperref}
\usepackage{multirow}
\usepackage{float}
\usepackage{lineno}
\usepackage{xspace}
\usepackage{subcaption}
\usepackage{booktabs}
\usepackage[usenames,dvipsnames]{color}
\usepackage{hyperref}
\usepackage{xspace}
\usepackage{multirow}
\usepackage{ragged2e}
\newcommand{\sqsn}{$\sqrt{s_{_{\rm{NN}}}} =$\xspace}

\newcommand{\pt}{$p_{\rm{T}}$\xspace}

\newcommand{\pd}{$p\mathrm{-}d$\xspace}
\newcommand{\dd}{$d\mathrm{-}d$\xspace}
\newcommand{\pH}{$p\mathrm{-}t$\xspace}
\newcommand{\pHe}{$p\mathrm{-}^{3}He$\xspace}

\begin{document}

\title{Probing coalescence of light nuclei via femtoscopy and azimuthal anisotropies}
\author{Yoshini Bailung$^{1}$} \thanks{yoshini.bailung.1@gmail.com}
\author{Sudhir Pandurang Rode$^{2}$}
\thanks{sudhirrode11@gmail.com}
\author{Neha Shah$^{3}$}
\thanks{nehashah@iitp.ac.in}
\author{Ankhi Roy$^{1}$}
\thanks{ankhi@iiti.ac.in}
\affiliation{$^{1}$Department of Physics, Indian Institute of Technology Indore, Simrol, Indore, Madhya Pradesh, India}
\affiliation{$^{2}$Veksler and Baldin Laboratory of High Energy Physics, Joint Institute for Nuclear Research, Dubna 141980, Moscow region, Russian Federation}
\affiliation{$^{3}$Department of Physics, Indian Institute of Technology Patna, Bihta, Patna, Bihar, India}

\begin{abstract}
The production mechanism of light nuclei in heavy-ion collisions is vital to understanding the intricate details of nucleon-nucleon interactions. The coalescence of nucleons is a well-known mechanism that attempts to explain the production mechanism of these light clusters. This work investigates the formation mechanism of these nucleon clusters with a combination of coalescence and femtoscopy of nucleons and nuclei. It is achieved by appending a coalescence and correlation afterburner (\texttt{CRAB}) to the \texttt{SMASH} transport model. To have a proper view of the anisotropy of light nuclei clusters, a mean-field approach to \texttt{SMASH} is applied. The anisotropic coefficients of various light nuclei clusters are calculated and compared to experimental measurements. To incorporate hydrodynamics into the picture, the anisotropic measurements are completed in a hybrid \texttt{SMASH}+\texttt{vHLLE} mode. In both approaches, the femtoscopy of nucleons and light nuclei is performed, reported with CRAB, and compared to the latest experimental measurements. An insight into cluster formation time is drawn by extracting the emission source size with the Lednick\'y-Lyuboshits (LL) model.
\end{abstract}
\date{\today}
\maketitle

\section{Introduction}
\label{introduction}
The light nuclei production is believed to be sensitive to the baryon density fluctuations and can be utilized to probe the QCD phase transition~\cite{STAR:2021ozh,NA49:2016qvu,Sun:2017xrx,Bastian:2016xna}. However, their production mechanism in heavy-ion collisions is not completely understood, despite various theoretical attempts made by an ensemble formalism via statistical hadronization, from nucleons' coalescence or stochastic processes~\citep{Andronic:2010qu,Cleymans:2011pe,Hillmann:2021zgj,Steinheimer:2012tb,Shah:2015oha,Zhao:2021dka,Glassel:2021rod,Sun:2020uoj,Dal:2015sha,Bailung:2023dpv,JETSCAPE:2022cob}. Existing and upcoming experimental measurements of light nuclei can provide a vast testing ground for these models in search of the nature of such low-energy QCD interactions.  Moreover, due to its bound nature, light nuclei states are an ideal probe for performing two-particle correlations and femtoscopy. Particle femtoscopy in heavy-ion collisions provides insight into particle emission's space-time geometry and their interactions~\cite{Lisa:2005dd,STAR:2014shf,ALICE:2020mfd}. Although femtoscopy for various hadrons has been measured, there is less exploration in the light nuclei regime. The ALICE collaboration first moved by looking into the \pd system and opening avenues to explore the emission of a few nucleon clusters~\citep{Singh:2022qmg,Appelshauser:2022ydz,ALICE:2023bny}. Understanding the interaction of these many-nucleon systems can assist in the realization of the intricacies of nucleon-nucleon interactions, as well as the equation of state (EoS) of neutron stars. Theoretical predictions have provided the possibility of having three (\pd) and even four (\dd, \pH, \pHe) nucleon-bound states that could be very well observed in heavy-ion collisions~\cite{Kievsky:2003eu,Sekiguchi:2019xvh,Pieper:2001ap,FOPI:1999emg,Boal:1990yh,Wang:2022hja,Jennings:1985km}. Recent efforts have been made by HADES~\cite{Stefaniak:2024fkf} and STAR~\cite{Mi:2022zig,Mallick:2023pcx} collaborations on the momentum correlations and femtoscopy of these states, providing an opportunity to understand their production mechanisms. There are two popular light nuclei production mechanisms, namely, nucleon coalescence and thermal emission, which are being pursued by the high-energy physics community. However, the formal approach seems more convincing, as the latter poses $``$ice in a fire'' puzzle.\\

In principle, the few-nucleon cluster production can be explored by combining femtoscopy and coalescence. Recently, STAR collaboration carried out such measurements in Au+Au collisions at \sqsn 3 GeV for \pd and \dd states, followed by a transport model calculation with \texttt{SMASH}~\cite{Mi:2022zig,SMASH:2016zqf}. The deuterons from coalescence have shown an excellent ability to describe the experimental measurements. There has been an inclination towards coalescence as a primary production mechanism of light nuclei. The emission source of the nucleon/nuclei pairs from the femtoscopic correlations shows a decreasing trend with the centrality in Au+Au collisions~\cite{Mi:2022zig}. This is due to the source sizes being non-trivially affected by the colliding matters' expansion time, radially and longitudinally. The measurement of emission space-time geometry of particles can assist in drawing a clearer description of the collective nature of expansion. This is of particular interest to understanding anisotropic flow in non-central heavy-ion collisions. The convolution between these two aspects of light nuclei production brings together a cohesive point of discussion about their production mechanism. Microscopic transport models such as \texttt{UrQMD}~\cite{Bass:1998ca}, \texttt{AMPT}~\cite{Lin:2004en}, and more recent \texttt{SMASH}~\cite{SMASH:2016zqf} are developed with one of many goals to interpret the generated anisotropies in heavy-ion collisions. Non-central heavy-ion collisions lead to azimuthally asymmetric emission of particles, the distribution of which can be decomposed into a Fourier expansion in the particles' azimuthal angle ($\phi$). Each Fourier harmonic constitutes a flow component, such as directed flow $v_{1} = \langle \cos \phi\rangle$ and elliptic flow $v_{2} = \langle \cos 2\phi\rangle$ corresponding to first and second harmonics respectively~\cite{Poskanzer:1998yz,STAR:2004jwm}.\\

Experimental evidence suggests that protons' elliptic flow coefficient is below zero at center-of-mass energies around and below 3 GeV~\cite{E895:1999ldn}. Depending on the time scale of expansion in low energy collisions, the expanding matter can be blocked by the spectators in the reaction plane~\cite{Danielewicz:2002pu}. Subsequently, this leads to a $``$squeeze-out'' effect if the blockage time from the spectators is more, forcing the emission out-of-plane~\cite{Danielewicz:1998vz,Nara:2021fuu}. This $``$effect'' is arguably connected to a softening or hardening of the EoS, which the models exploit to explain the experimental measurements~\cite{FOPI:2011aa,Petersen:2006vm,Hillmann:2018nmd,Konchakovski:2014gda}. Fluid dynamical treatment with a $`$soft' EoS fails to explain the negative elliptic flow. Here, the importance of mean-field interactions could be drawn out, which, in addition to the EoS, is an established approach to explain the flow of protons at low energies~\cite{Aichelin:1987ti,Isse:2005nk,LeFevre:2016vpp,Nara:2016hbg,Nara:2021fuu}. Many transport models can achieve this by parametrizing the Skyrme mean-field potential between the nucleons based on the $``$stiffness'' of the EoS. Calculations with \texttt{SMASH} and \texttt{UrQMD} at low energies have reported success in reproducing experimental data from HADES and STAR of light nuclei flow with the application of a hard EoS~\cite{Hillmann:2019wlt,Mohs:2020awg}.\\

In this work, we simulate Au+Au collisions at \sqsn 3 GeV using \texttt{SMASH} simulation package~\cite{SMASH:2016zqf}. Our choice of beam energy is influenced by a couple of reasons, the first being the availability of the experimental measurements of light nuclei observables at this beam energy by the STAR collaboration\cite{STAR:2021ozh,Mi:2022zig}. Moreover, this beam energy falls into the region where new measurements from upcoming experiments such as Compressed Baryonic Matter (CBM) at FAIR~\cite{Almaalol:2022xwv} and Multi-Purpose Detector (MPD) at NICA~\cite{Golovatyuk:2016zps} are foreseen. We test our $``$in-situ'' coalescence model on different cluster formation times. This transport model is also used to generate the phase-space information of the nucleons at different freezeout times. For closure, we employ a hard EoS mean field parametrization of the Skyrme potential in \texttt{SMASH}, as well as a hybrid treatment, where a fluid dynamical evolution is carried out with the \texttt{vHLLE} package~\cite{Karpenko:2013wva,Schafer:2021csj}. We perform femtoscopic correlations on these clusters to validate the impact of different cluster formation times. The correlation afterburner (\texttt{CRAB}) model~\cite{Pratt:1997cb} is used to generate the momentum correlations of \pd and \dd pairs for varying centralities in Au+Au collisions at \sqsn 3 GeV, which is compared to experimental results. Following the momentum correlations, we extract the source sizes of the light cluster systems, by fitting the Lednick\'y Lyuboshits (LL) model~\cite{Lednicky:1981su}, with a Gaussian ansatz for the source. By varying the freeze-out times of the nucleons, we gain insight into the formation times from the emission sizes of these clusters. The freeze-out time is expected to be different for different nucleon bound states, which is tested out for \pd, and \dd states. The results are also supplemented with the directed and elliptic flow of light clusters in mid-central Au+Au collisions.\\

The paper is arranged in the following manner: Section~\ref{smashcoal} is divided into subsections~\ref{smash} and~\ref{coal} for the event generation methodology with \texttt{SMASH} and the coalescence model, respectively. Likewise, Section~\ref{femtocrab} has subsections~\ref{femto} and~\ref{crab} dedicated to femtoscopy, explaining the source size estimation with the LL model and CRAB for the momentum correlations. Section~\ref{results} presents the results, the correlation functions shown in subsection~\ref{corrfuncs}, and the anisotropic flow calculations in subsection~\ref{flows}. We conclude the paper by summarising the findings in Section~\ref{conclusions}.

\section{\texttt{SMASH} and Coalescence}\label{smashcoal}
\subsection{The \texttt{SMASH}  transport model}\label{smash}
The hadronic transport model \texttt{SMASH}, or $``$Simulating Many Accelerated Strongly interacting Hadrons" is a many-interaction model developed to extend the exploration of the QCD phase diagram to the low and intermediate beam energies~\cite{SMASH:2016zqf}. In this regime, the prevailing degrees of freedom are that of hadrons, which usually do not require a hybrid treatment, unlike the dynamical transport models, which impose hybrid treatment of each stage of evolution of the collision. These dynamical models face challenges in the intermediate beam energies, which are of great interest to unlocking the attributes of the QCD phase diagram. However, \texttt{SMASH} has also been designed to operate in a hybrid mode, which allows it to be coupled with a relativistic hydrodynamic expansion. The hybrid \texttt{SMASH} is coupled to the \texttt{vHLLE}; a 3+1D viscous hydrodynamic model~\cite{Karpenko:2013wva} that follows a well-known fluid expansion where the level of dilution is examined by a critical energy density cutoff. For the initial and final stages of the evolution, \texttt{SMASH} operates in a pure transport mode, following hydrodynamic expansion by the hadron sampler and a hadronic afterburner. Although at very low energies, a hydrodynamic treatment of the collision is not essential, it was an obvious choice in the light nuclei aspect to show a better description of the deuteron yields with the hydro turned on than a pure transport \texttt{SMASH}. A well-detailed description of the modules can be found in the reference~\cite{Schafer:2021csj}.\\

For the mean-field approach, the interaction potentials in \texttt{SMASH} are parameterized to replicate the stiffness of EoS. The Skyrme and symmetry potential in \texttt{SMASH} is expressed as
\begin{equation}
    U = U_{\mathrm{sk}} + U_{\mathrm{sym}}
\end{equation}
which, individually, are written as
\begin{eqnarray}
    U_{\mathrm{sk}} &=& A\left(\frac{\rho_{B}}{\rho_{0}}\right) + B\left(\frac{\rho_{B}}{\rho_{0}}\right)^{\tau}\\
    U_{\mathrm{sym}} &=& \pm 2 S_{\mathrm{pot}}\frac{\rho_{I_{3}}}{\rho_{0}}
\end{eqnarray}
where $\rho_{B}$ is the net baryon density, $S_{\mathrm{pot}}$ is the symmetry potential energy, $\rho_{0} = 0.168~\mathrm{fm^{-3}}$ is the nuclear ground state density, $\rho_{I_{3}}$ is the density of the relative isospin projection. A hard EoS corresponds to a larger value of compressibility ($K$), which is tuned via the parameters $A,~B,~\mathrm{and}~\tau$, the values for which are studied and used in references~\citep{Hillmann:2019wlt,Mohs:2020awg,Kruse:1985hy}.

\subsection{The Coalescence Afterburner}
\label{coal}
With the phase-space information of the nucleons, a coalescence approach is applied to produce nuclei out of closely placed nucleons. This work takes a Wigner coalescence approach, where the nuclei wave function is expressed as a Wigner probability density~\cite{Bailung:2024sca}. Coalescence is achieved by checking the level of overlap between the phase space of the nucleons and the nuclei's Wigner probability. The invariant yields for deuteron and triton/helium-3 can be expressed in the form
  \begin{eqnarray}
  \label{maineqn2}
  &\frac{d^3N_{d(t/^{3}\mathrm{He})}}{dp_{d(t/^{3}\mathrm{He})}^3} = \mathcal{S}_{d(t/^{3}\mathrm{He})}\int d^3\mathbf{r}_{d(t/^{3}\mathrm{He})}d^3\mathbf{r}(\mathbf{r}^{\prime})d^3\mathbf{q}(\mathbf{q}^{\prime})\\ \nonumber
  &\times\mathcal{D}(\mathbf{r}(\mathbf{r^{\prime}}), \mathbf{q}( \mathbf{q^{\prime}}))\\ \nonumber
  &\times W_{pn(p/n-np)}\left(\frac{\mathbf{p}_{d(t/^{3}\mathrm{He})}}{2} \pm \mathbf{q(q^{\prime})},
  \mathbf{r}_{d(t/^{3}\mathrm{He})} \pm \frac{\mathbf{r(r^{\prime})}}{2}\right)
\end{eqnarray}
respectively, where $\mathbf{r}_{d(t/^{3}\mathbf{He})},~\mathbf{p}_{d(t/^{3}\mathbf{He})}$, and $\mathbf{r(r^{\prime})},~\mathbf{q(q^{\prime})}$ are the cluster spatio-momenta and the  respectively. $\mathcal{S}_{d(t/^{3}\mathrm{He})}$ is the statistical spin-isospin averaging factor, $W_{pn(p/n-np)}$ is the nucleon pair selection probability term, and $\mathcal{D}$ is the Wigner probability density. The Wigner density can be derived from the light-nuclei structure, expressed by its quantum mechanical wave function, and can be related to the source function of the two and three-nucleon clusters. The deuteron wave function is approximated to be a single Gaussian function $\varphi (r) = \frac{1}{(\pi \sigma_{d}^2)^{3/4}} e^{-\frac{r^2}{2\sigma_{d}^2}}$ with rms charge radius $\sigma_{d}$ = 2.14 fm\cite{Bellini:2018epz}; consequently the Wigner probability density becoming
\begin{equation}
    \mathcal{D}(\mathbf{r}, \mathbf{q}) = 8e^{-\frac{\mathbf{r}^2}{\sigma_{d}^2} - \mathbf{q}^2 \sigma_{d}^2}.
\end{equation}
For the three-nucleon clusters ($t$, $^{3}\mathrm{He}$), the Wigner probability is modified as

\begin{eqnarray}
    \mathcal{D}(\mathbf{r}, \mathbf{q}, \mathbf{r^{\prime}}, \mathbf{q^{\prime}}) = 8^{2}e^{-\frac{\mathbf{r}^2}{\sigma_{t/^{3}\mathrm{He}}^2} - \mathbf{q}^2 \sigma_{t/^{3}\mathrm{He}}^2} e^{-\frac{\mathbf{r}^{\prime 2}}{\sigma_{t/^{3}\mathrm{He}}^2} - \mathbf{q}^{\prime 2} \sigma_{t/^{3}\mathrm{He}}^2}.
\end{eqnarray}
where $\sigma_{t} = $ 1.75 fm, and $\sigma_{^{3}\mathrm{He}} = $ 1.96 fm, are the charge rms radius of the triton and helium-3 respectively~\cite{Bellini:2018epz}. 

\begin{figure*}
    \centering
\includegraphics[scale = 0.8]{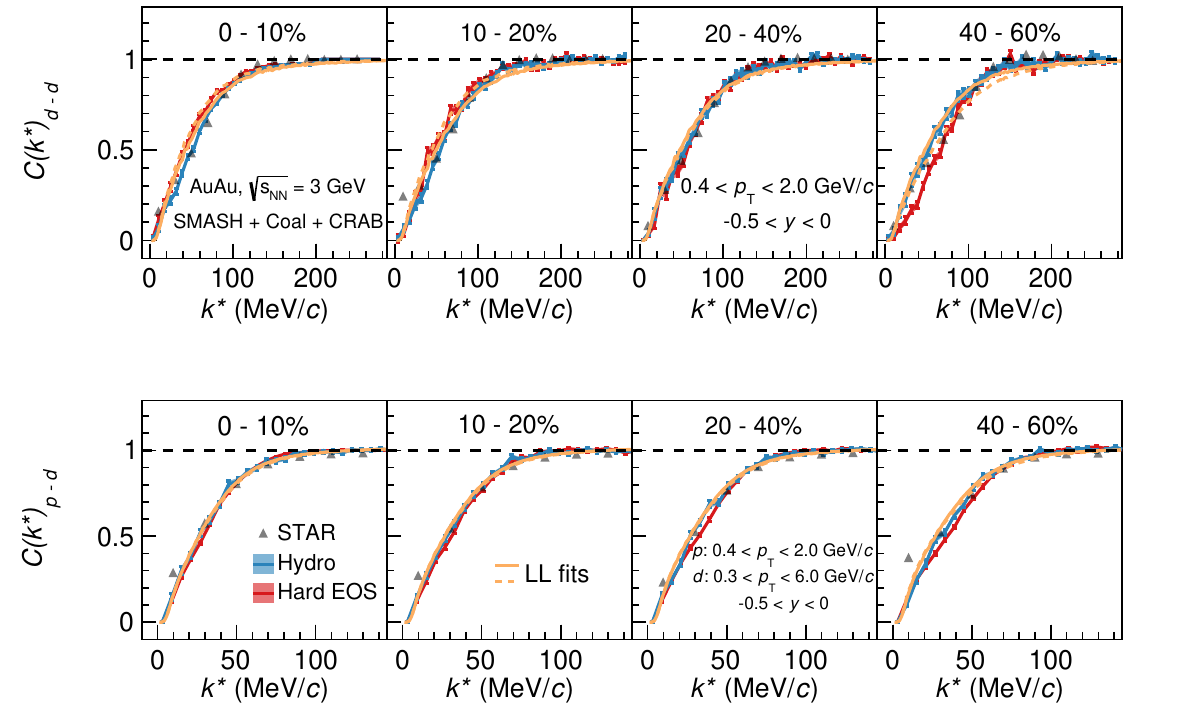}
\caption{\justifying Momentum correlation of \pd (bottom) and \dd (top) pairs in Au+Au collisions at \sqsn 3 GeV in different centralities from \texttt{SMASH} + Coal + \texttt{CRAB}, compared to STAR preliminaries~\cite{Mi:2022zig}. Predictions are presented for hydro and hard EoS \texttt{SMASH} fitted with the LL model shown in solid and dashed red lines, respectively.}
\label{fig:corr}
\end{figure*}

\section{Femtoscopy and CRAB}\label{femtocrab}
\subsection{Femtoscopy of light nuclei}\label{femto}
Femtoscopy is performed via two-particle momentum correlations of like or unlike particle pairs, which gives the space-time picture of their emission from the interaction region~\cite{Lisa:2005dd,STAR:2014shf,ALICE:2020mfd}. Experimentally, the correlation function for a two-nucleon system can be obtained by taking the relative momentum ($q^*$) in the pair-rest frame of the nucleons. It can be defined as 
\begin{equation}
    C(q^*) = \frac{1}{\mathcal{N}}\frac{A(q^*)}{B(q^*)}
    \label{Cexp}
\end{equation}
where $A(q^*)$ and $B(q^*)$ are the two-nucleon correlation distributions in the same and mixed event, respectively, and $\mathcal{N}$ is a normalizing constant. The same distribution, in theory, can be expressed as a convolution of the source function corresponding to the emission system and the wave function or the Bethe-Salpeter amplitude ($\Psi$) of the two-nucleon system. The source $S(\mathbf{r})$ can be estimated as
\begin{equation}
    C(q^{*}) = \int d^3r S(\mathbf{r}) \mid\Psi(\mathbf{r},q^{*})\mid^{2}
    \label{Ctheo}
\end{equation}
For identical pairs, the source function can be assumed to be a Gaussian of source radius $r_{0}$, to yield the correlation function of correlation strength $\lambda$ as described below
\begin{equation}
\label{simplesource}
    C(q^{*}) = 1 + \lambda e^{-q^{*2}r_{0}^{2}}
\end{equation}
In this study, however, we use the Lednick\'{y}-Lyuboshits (LL) model~\cite{Lednicky:1981su,Morita:2019rph}, which is more exhaustive to Equation~\ref{simplesource} that considers final state interactions (FSIs) in the form of Coulomb and strong interactions, and quantum statistical fluctuations between the nucleons. The full correlation function for identical pairs can be defined as
\begin{eqnarray}
    C(q^{*}) = 1 + \lambda\left(e^{-q^{*2}r_{0}^{2}} + \frac{1}{2}\left(\left|\frac{f^{S}_{c}(k^{*})}{r_{0}}\right|^{2} + \frac{4\mathfrak{R}f^{S}_{c}(k^{*})}{\sqrt{\pi}r_{0}}\right. \right. \nonumber \\
    \left. \left. \times F_{1}(q^{*}r_{0}) - \frac{2\mathfrak{I}f^{S}_{c}(k^{*})}{\sqrt{\pi}}F_{2}(q^{*}r_{0})\right)\right).
\end{eqnarray}
For non-identical pairs, the same expression can be approximated to

\begin{eqnarray}
    C(q^{*}) \approx 1 + \frac{1}{2}\left(\left|\frac{f^{S}_{c}(k^{*})}{r_{0}}\right|^{2} + \frac{4\mathfrak{R}f^{S}_{c}(k^{*})}{\sqrt{\pi}r_{0}}\right. \nonumber \\
    \left. \times F_{1}(q^{*}r_{0}) - \frac{2\mathfrak{I}f^{S}_{c}(k^{*})}{\sqrt{\pi}}F_{2}(q^{*}r_{0})\right).
\end{eqnarray}
$f^{S}_{c}(k^{*})$ is the forward scattering amplitude with Coulomb interactions for a given total spin $S$, and functions $F_{1}(z) = \frac{1}{2} e^{-z^{2}}\int_{0}^{z}e^{x^{2}} dx$, and $F_{2}(z) = \frac{1}{z}(1-e^{-z^2})$. The LL model has been extensively used to extract source sizes and interaction parameters in experimental femtoscopy studies for various particle systems. Light nucleon/nuclei correlations of the \pd and \dd kind have been studied by the STAR collaboration for Au+Au collisions at \sqsn 3 GeV across different centralities~\cite{Mi:2022zig}. This study will use the LL model to fit the nuclei correlation distributions obtained via \texttt{CRAB}, to which the light-nuclei states from the coalescence model will be fed.
\subsection{Correlation afterburner (CRAB)}\label{crab}
To perform the momentum correlation of the nucleon/nuclei states, with the inclusion of quantum statistics and FSIs, a $``$correlation-afterburner'' \texttt{CRAB}~\cite{Pratt:1997cb} is coupled to \texttt{SMASH}. To perform light nuclei femtoscopy, the coalescence model generates the input to \texttt{CRAB}, to carry out momentum correlations for \pd and \dd states. The working of \texttt{CRAB} is based on the correlation function defined as
\begin{equation}
    C(p, q) = \frac{\int d^{4}x_{1}d^{4}x_{2}S_{1}(x_{1},p_{2})\cdot S_{2}(x_{2},p_{1}) |\phi(q, r^{*})|^{2}}{\int d^{4}x_{1}S_{1}(x_{1},p_{1}) \int d^{4}x_{2}S_{2}(x_{2},p_{2})}
\end{equation},
where $(x,p)$ is the phase space information of the nucleons/nuclei at freeze-out, $S_{i}(x,p)$ is the probability of emitting particle $i$ at phase space $(x_{i},p_{i})$, $p(q)$ are the total (relative) momentum, $r^{*}$ is the relative position, and $\phi(q, r^{*})$ is the overlap wave function of the pair; used to build the pair correlation functions. We compare the results to the experimental measurements with the correlations generated from the combination of \texttt{SMASH}, coalescence, and \texttt{CRAB}.

\begin{figure}
    \centering
    \includegraphics[scale = 0.4]{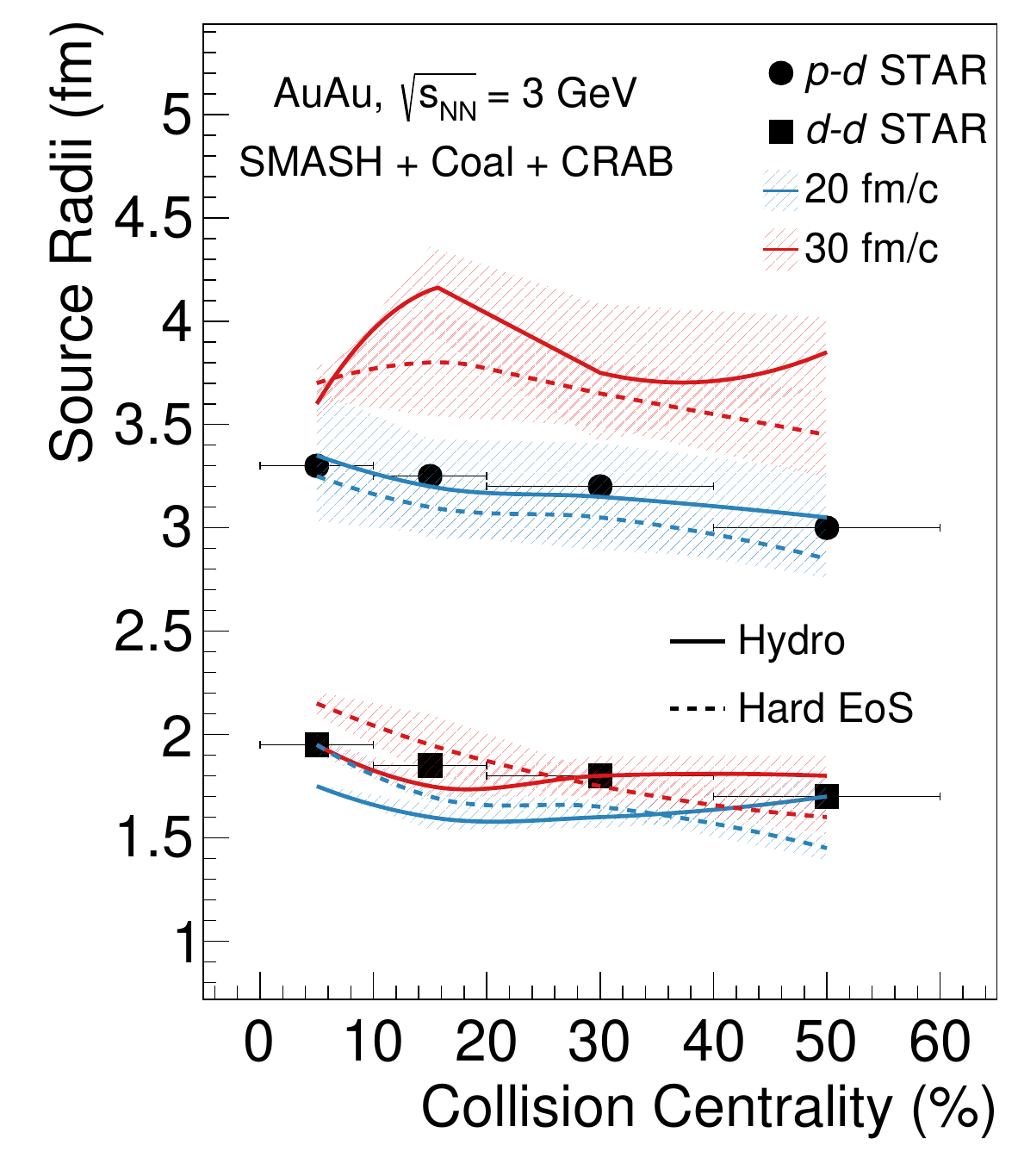}
    \caption{\justifying Source radius extracted from LL fits to \pd and \dd correlation functions from STAR prelimaries~\cite{Mi:2022zig}, and from two cluster formation times in hydro (solid lines) and hard EoS (dashed lines) \texttt{SMASH} + Coal.}
    \label{fig:radiusvscent}
\end{figure}

\section{Results and Discussion}
\label{results}
  
\subsection{Correlation Functions}
\label{corrfuncs}

Figure~\ref{fig:corr} shows the predictions of momentum correlation functions from \texttt{SMASH} + Coal + \texttt{CRAB} fitted to the LL model, of \pd and \dd pairs in different collision centralities. \texttt{SMASH} is configured in two ways viz., with hydro and with a hard EoS. The correlation distributions are obtained from the \texttt{CRAB} afterburner, which takes in the nucleon (nuclei) phase-space from \texttt{SMASH} (\texttt{SMASH} + Coal). As \texttt{CRAB} is not designed for \pd correlations, the potential for \pd interactions is adapted from~\cite{Jennings:1985km}. The correlations are tested on two cluster formation times 20 and 30 fm/c up to which \texttt{SMASH} hadronic cascade is evolved and at which point the coalescence of nucleon pairs is carried out. The \pd and \dd femtoscopy results presented in Figure~\ref{fig:corr} are for cluster formation times 20 fm/c and 30 fm/c, respectively, for hydro and hard EoS \texttt{SMASH}. The results are compared to the preliminary measurements from STAR~\cite{Mi:2022zig}. Except at 40-60\% centrality, where a deviation between the hydro and hard EoS is noticed, the results show excellent agreement with the data for \pd and \dd correlations. A closer look reveals that the hydro predictions are slightly better than the hard EoS selection, especially noticeable for \dd pairs at peripheral collisions.\\

To gain a more quantitative insight and compare the formation times, we extract the source sizes of the \pd and \dd pairs. The source sizes are extracted by fitting the LL model to the correlation distributions, assuming a Gaussian emission source. The fits are performed on the \texttt{SMASH} + Coal + \texttt{CRAB} results, as well as on the STAR preliminary measurements~\cite{Mi:2022zig}. In Figure~\ref{fig:corr}, we report the fits to the \texttt{SMASH} + Coal + \texttt{CRAB} outputs across different centralities for \pd and \dd correlation functions for hydro and hard EoS configurations. The source sizes reported in Figure~\ref{fig:radiusvscent} show a decreasing dependence with centrality for the STAR data. The \pd cluster source sizes are larger, lying between 3 -- 3.5 fm, in contrast to  1.8 -- 2 fm for the \dd system. The \texttt{SMASH} + Coal + \texttt{CRAB} results for \pd pairs are well reproduced at 20 fm/c for hydro, with a slight underestimation by the hard EoS. At 30 fm/c, the source sizes overestimate the data, showing deviations upto~30\%. The results for \dd pairs also predict the decreasing trend of source sizes with centrality, affirmed by the experimental results. Here, the scale of disagreement between the cluster formation times is less compared to the \pd pairs. Hydro \texttt{SMASH} offers the closest description at 30 fm/c. The hard EoS show~10\% overestimation for central collisions, with closer description at mid-central to peripheral collisions. At 20 fm/c, the source radii are slightly lower than 30 fm/c, except at 40-60\% for hydro.\\

Even with finely spaced cluster formation times, we see variation in the \pd and \dd correlations reflected in the emission source radii. Based on the agreement with the data, we report that a cluster time at 30 fm/c for the deuteron is closer to describing the emission of \dd pairs, whereas a shorter time between 20--30 fm/c is preferred for the proton and deuteron pairs for the \pd cluster. This noteworthy observation indicates the formation time of differently populated baryon clusters formed in these heavy-ion collisions.

\begin{figure}[h]
    \centering
    \includegraphics[scale=0.5]{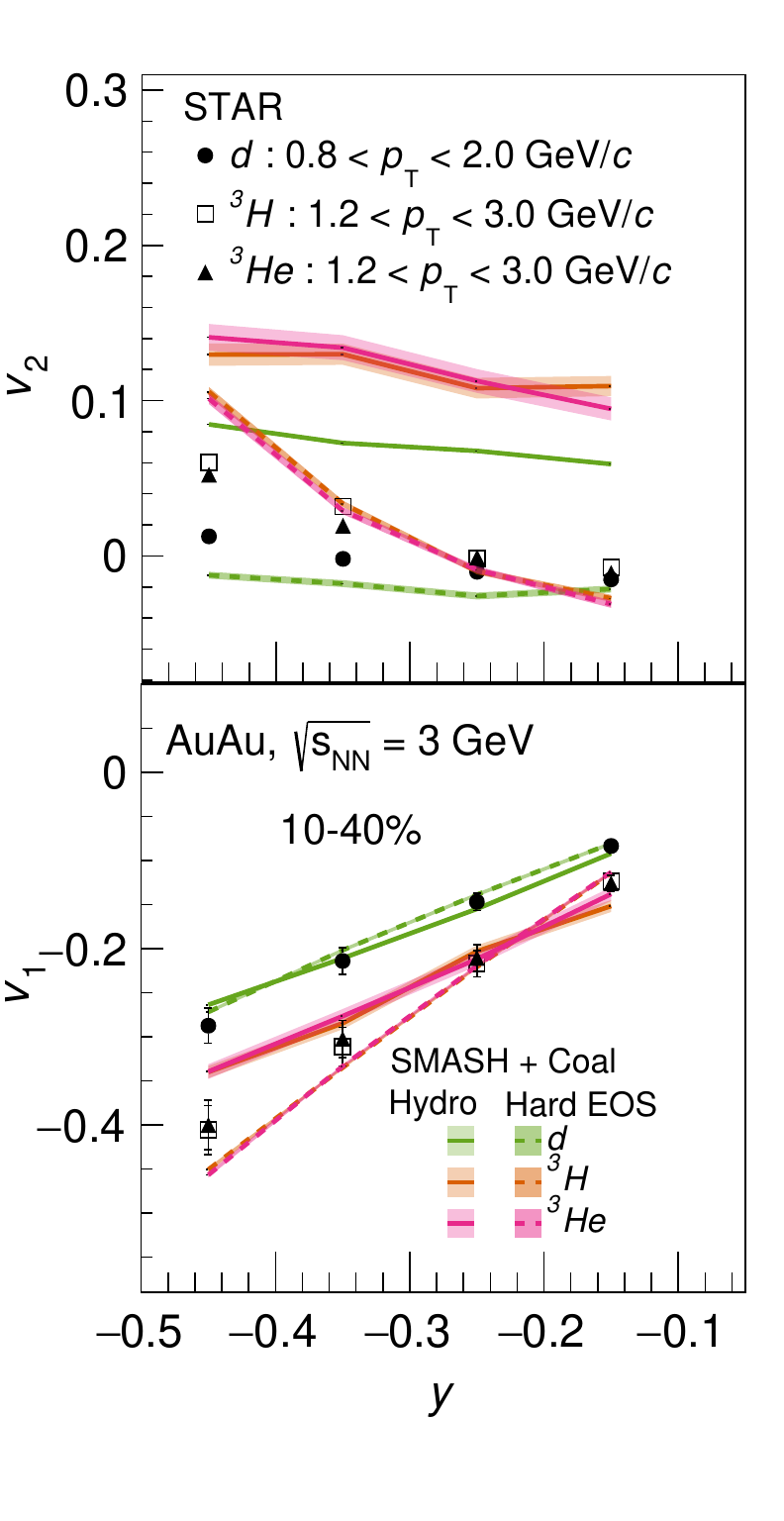}
    \caption{\justifying Elliptic flow ($v_2$) and directed flow ($v_{1}$) of deuterons, tritons and helium-3 produced in non-central Au+Au collisions as function of rapidity using \texttt{SMASH} + Coal. Results are compared to experimental measurements from STAR~\cite{STAR:2021ozh}.}
    \label{fig:v1v2vsy}
\end{figure}

\begin{figure}[h]
    \centering
    \includegraphics[scale=0.4]{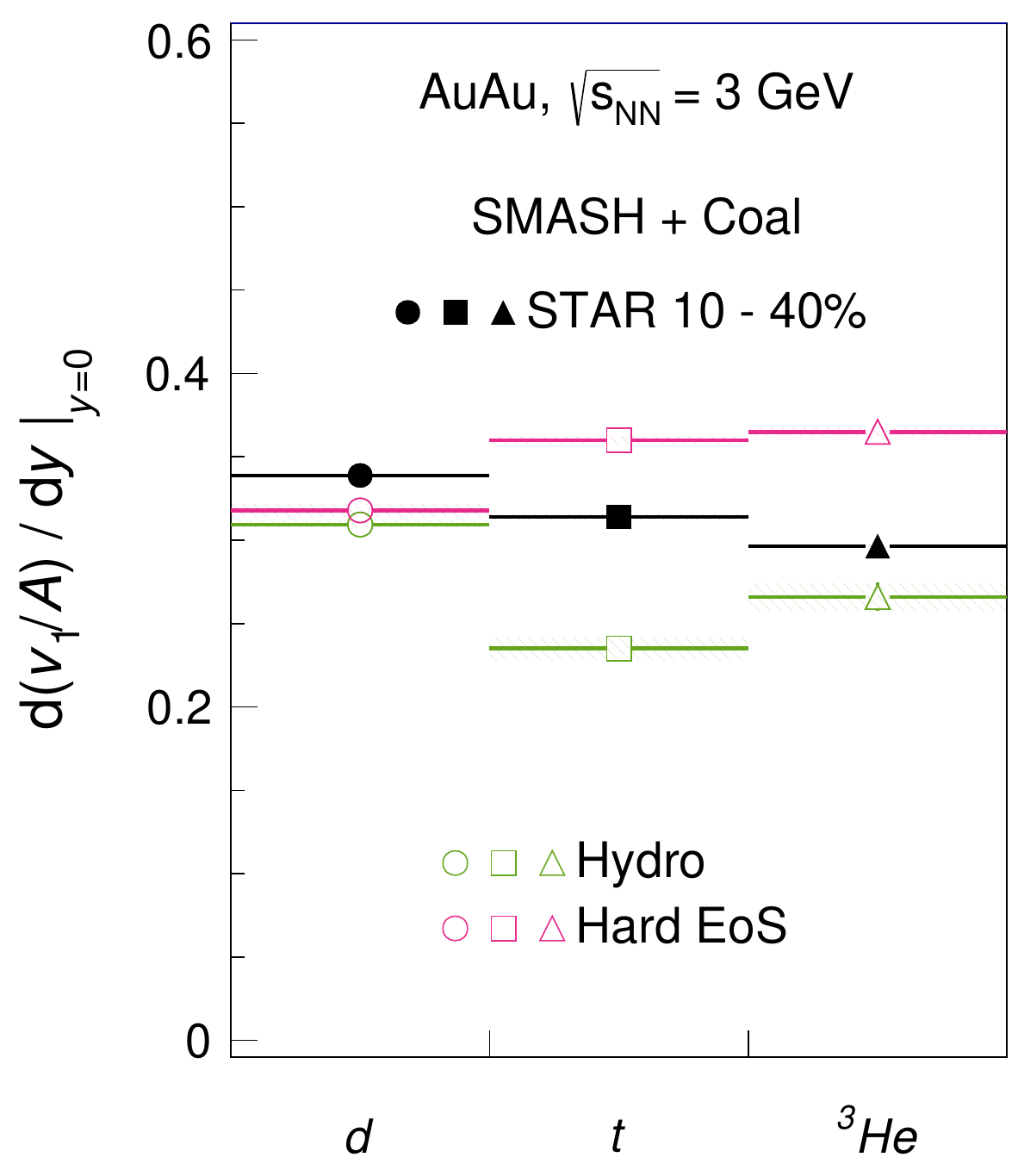}
    \caption{\justifying $dv_{1}/dy$ of deuterons, tritons and helium-3 produced in non-central Au+Au collisions using \texttt{SMASH} + Coal compared to experimental measurements from STAR~\cite{STAR:2021ozh}.}
    \label{fig:slopeplot}
\end{figure}

\begin{figure*}
    \centering
    \includegraphics[scale=0.7]{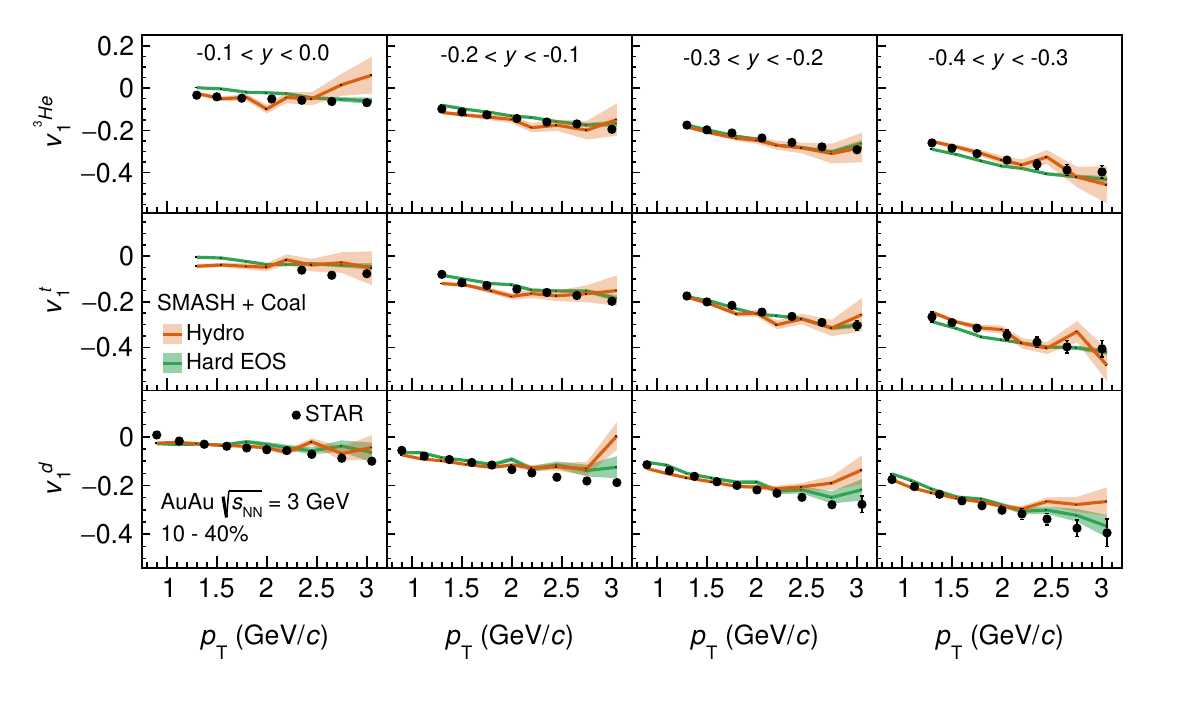}
    \caption{\justifying Directed flow ($v_1$) as function of \pt of deuterons, tritons and helium-3 produced in non-central Au+Au collisions using \texttt{SMASH} and their comparison with experimental measurements from STAR experiment~\cite{STAR:2021ozh}.}
    \label{fig:v1vspt}
\end{figure*}

\begin{figure*}
    \centering
    \includegraphics[scale=0.7]{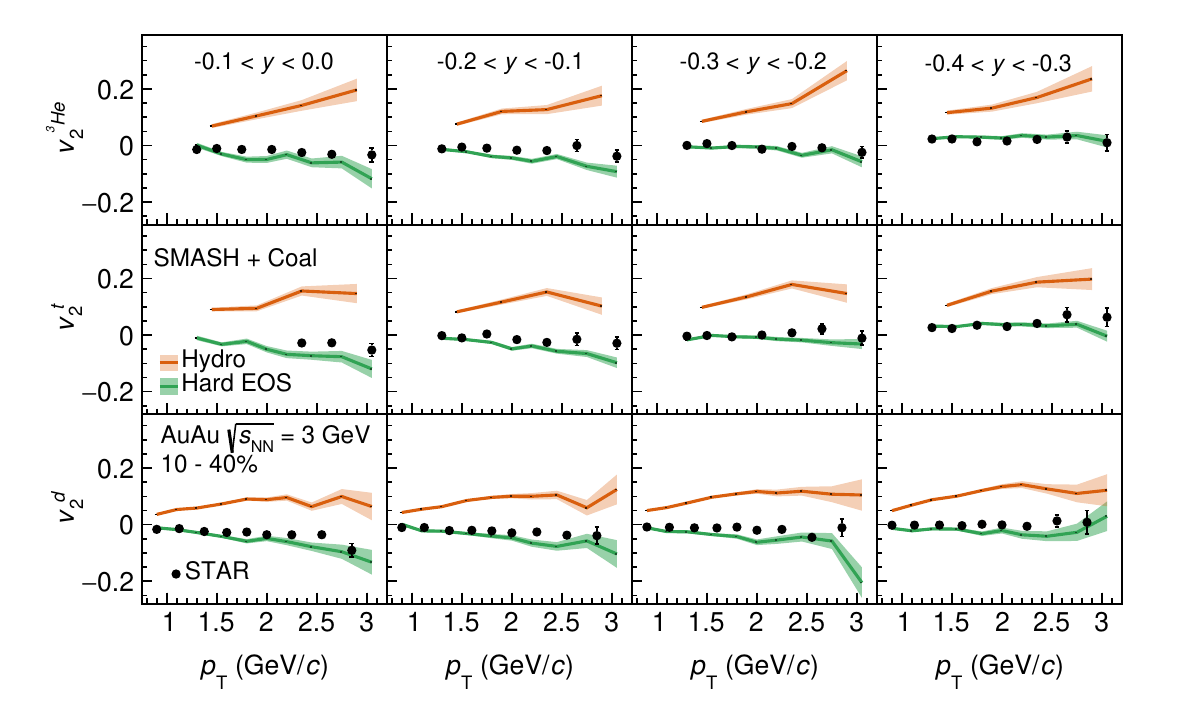}
    \caption{\justifying Elliptic flow ($v_2$) as function of \pt of deuterons, tritons and helium-3 produced in non-central Au+Au collisions using \texttt{SMASH} and their comparison with experimental measurements from STAR experiment~\cite{STAR:2021ozh}.}
    \label{fig:v2vspt}
\end{figure*}

\subsection{Anisotropic Flow}
\label{flows}

Now, we focus on estimating the azimuthal anisotropies of light nuclei in 10-40\% central Au+Au collisions at \sqsn 3 GeV. We estimate the directed ($v_{1}$) and elliptic flow ($v_{2}$) of deuterons, triton, and helium-3 and compare it with the extensive set of measurements performed by STAR collaboration~\cite{STAR:2021ozh}. Like the previous subsection, the flow coefficients are estimated for two variants of \texttt{SMASH}. The freeze-out time is taken from comparing the correlation functions from the previous subsection. The kinematic selections are kept the same as reported by the STAR measurements. We begin by studying the rapidity dependence of these flow coefficients, especially $v_{1}$, as it can reveal insights into the longitudinal dynamics of the collision. Figure~\ref{fig:v1v2vsy} presents the dependence of $v_{1}$ and $v_{2}$ of light nuclei as a function of rapidity ($y$), in specific \pt regions ($0.8<$ \pt $<2.0$ GeV/$c$ for deuteron and $1.2<$ \pt $<3.0$ GeV/$c$ for triton/helium-3).  Here, the coefficients of all species show considerable sensitivity to different \texttt{SMASH} modes, in particular, $v_{2}$. The hydro \texttt{SMASH} mode overestimates the measured $v_{2}$ of all light nuclei species by almost factor 2.
 
In addition to this and, more importantly, the \texttt{SMASH} + Coal + hydro variant fails to describe the negative values of $v_{2}$ (out-of-plane flow) at central rapidity intervals. It is worth noting that the hydro mode appears to reproduce the trend of $v_{2}$. On the contrary, \texttt{SMASH} + Coal + hard EoS reproduce the negative values and reasonably describe the light nuclei $v_{2}$ measurement at central rapidity; however, they slightly underestimate forward rapidities. This is possibly due to an effect of spectators or fragments from beam remnants that have an increasing rapidity dependence. An account of mean-field contributions seems to help describe the $v_{2}$ at such low beam energy. In retrospect, the $v_{1}$ of light nuclei is qualitatively described by both modes of \texttt{SMASH}. Hard EoS provides a closer description of the rapidity dependence of deuteron $v_{1}$, as compared to hydro, while for triton and helium-3, the modes under and overestimate towards increasing rapidity ($-0.4<y<-0.3$) respectively. However, both hydro mode and hard EoS reproduce the mass ordering in both $v_{1}$ and $v_{2}$ as observed in experimental measurement.\\

The slope of directed flow ($dv_{1}/dy$) is an interesting observable for the investigation since it is sensitive to the underlying EoS as well as one of the candidates to provide insights about the dynamics of the QCD medium. As mentioned above, both modes of \texttt{SMASH} describe the $v_{1}$ slope; however, we have performed quantitative comparison with experimental measurements.  In Figure~\ref{fig:slopeplot}, we present this observable of light nuclei with hydro and hard EoS \texttt{SMASH} and compare it with STAR measurements~\cite{STAR:2021ozh}. The slope is estimated by fitting the $v_{1}$ within the rapidity window, $-0.5<y<0.0$, using a first-order polynomial scaled to the respective mass numbers. The slopes corresponding to the deuterons from the hard EoS configuration are closer to the experimental results, showing about 6\% deviation, while the hydro results show a larger deviation $\sim$9\%. On the other hand, the slopes for triton and helium deviate on either side of the data, creating a band with hard EoS and hydro with deviation up to $\sim25\%$.\\

Moving forward, the transverse momentum dependence of directed flow coefficient ($v_{1}$) of $d,~t$, and $^{3}He$ as a function of \pt in different rapidity intervals is shown in Figure~\ref{fig:v1vspt}. Both \texttt{SMASH} modes, hard EoS, and hydro agree with the measurements across the whole \pt range from 1 to 3 GeV/$c$ and rapidity intervals. The \pt dependence of elliptic flow coefficients ($v_{2}$) of light nuclei in different rapidity regions are shown in Figure~\ref{fig:v2vspt}. The \texttt{SMASH} modes show deviations similar to the results in Figure~\ref{fig:v1v2vsy}, with a positive and increasing with \pt $v_2$ for  $d,~t$, and $^{3}He$ with hydro turned on. However, the failure of hydro in reproducing the sign suggests that \texttt{SMASH} + Coal + hydro at a low center of mass energies may not be enough to describe the $``$squeeze-out'' of colliding matter that eventually leads to a negative flow. The hard EoS reproduces this phenomenon well across \pt and rapidity region, asserting the importance of the mean-field approach. To check the robustness of the results, similar calculations of these flow coefficients were carried out using different freeze-out times, and it was found that it did not change the message of our findings. \\

In the literature~\cite{Danielewicz:2002pu} and~\cite{Mohs:2020awg}, it has been argued that at low energies, neither soft nor hard EoS describe the $v_{1}$ and $v_{2}$ together convincingly. However, we have seen in our work that, unlike hard EoS, soft EoS failed to explain the rapidity and \pt dependence of both $v_{1}$ and $v_{2}$. Since different transport model approaches have various underlying degrees of freedom, different conclusions regarding mean-field interaction strengths can be drawn. It has already been seen that existing FOPI data~\cite{FOPI:2011aa} is very nicely explained by soft EoS~\cite{Aichelin:1987ti}, and approaches with a hard EoS over \texttt{UrQMD} shows agreement with the NA49 and HADES measurements~\citep{Petersen:2006vm, Hillmann:2018nmd, Hillmann:2019wlt}. 

\section{Conclusions}\label{conclusions}
In this article, we investigate the momentum correlation function of nucleons and light nuclei and gain insight into their production times and their sensitivity to the initial anisotropy in heavy-ion collisions. The \texttt{SMASH} transport model generates the nucleon phase space, passing through a coalescence afterburner to form the light nuclei states. \texttt{SMASH} is operated in two configurations: a hybrid configuration with a viscous hydrodynamics treatment and a mean-field configuration with a hard EoS.\\

The protons and deuterons from coalescence are used as input to the \texttt{CRAB} afterburner to generate the momentum correlations of \pd and \dd states. The correlation functions are obtained at different collision centralities compared to experimental measurements at STAR. The correlations are fitted with the LL model, assuming a Gaussian emission source. The emission source radius of the \pd, \dd states from the fits shows a decreasing trend with increasing centrality. The excellent agreement of these predictions with the data asserts the importance of coalescence playing an important role in light nuclei formation. The calculations are checked by varying the cluster formation time of the nucleons/nuclei. We report that a difference of 10 fm/c is observed between the \pd cluster and the \dd cluster, the former being produced earlier. This is a testament to the sensitivity of formation times for different sources during the medium evolution.\\

To look into the influence of formation times on the collective expansion of the light clusters, we look at the directed and elliptic flow coefficients of the deuteron, helion, and helium-3 from coalescence for mid-central Au+Au collisions. In particular, we compare a mean-field potential approach to a hybrid approach of particle production in low-energy Au+Au collisions. Both treatments do not have noticeable differences for the directed flow of light nuclei but differ considerably for $v_{2}$, and do not show any dependence on the formation times of the clusters. The results allow us to conclude that a mean-field treatment with a hard EoS parametrization gives a precise description at low collision energies. We also reiterate that \texttt{SMASH} with soft EoS failed to describe rapidity and \pt dependence of both flow coefficients.\\

The findings shared in this analysis provide new avenues to extend and perform similar studies to other light and hypernuclei species ($p$--$^{3}He$, $p$--$t$, $t$--$t$, $p$--$_{\Lambda}^{3}H$). Femtoscopy studies of these states can provide more insight into the nucleon-nucleon interactions while estimating their formation times via coalescence. Moreover, the estimation of the emission source sizes scanned over a collision energy range would shed more light on the energy dependence and the formation time of these clusters. Such studies can be carried out once more data becomes available in the near future.\\

\section{Acknowledgements} 
Y.B. thanks all the members of the SMASH collaboration, Prof Scott Pratt for the \texttt{CRAB} package and Prof Lednick\'y for providing us with the LL model. Y.B. is thankful to Swapnesh Khade for the fruitful discussions. This work uses computational facilities supported by the DST-FIST scheme via SERB Grant No. SR/FST/PSI-225/2016, by the Department of Science and Technology (DST), Government of India.






 \end{document}